\documentclass{CUP-JNL-EXP}

\usepackage{latexsym}
\usepackage{graphicx}
\usepackage{multicol,multirow}
\usepackage{amsmath,amssymb,amsfonts}%%
\usepackage{mathrsfs}
\usepackage{amsthm}
\usepackage{apacite}
\usepackage{rotating}
\usepackage{appendix}
\usepackage[authoryear]{natbib}
\usepackage{mathtools}
\usepackage{amsmath}
\usepackage[percent]{overpic}

\articletype{Physics and Astronomy}
\articlesubtype{Novel result}
\jname{Experimental Results}
\jyear{2021}

\DeclareRobustCommand\sbseries{\fontseries{sb}\selectfont}
\DeclareTextFontCommand{\textsb}{\sbseries}

\begin{document}

\begin{Frontmatter}

\title[Shock ID]
{Shock identification and classification in 2D MHD compressible turbulence - Orszag-Tang vortex}

%\author[1,2]{Hiroshi Sakai}
\author*[1]{B. Snow}\email{b.snow@exeter.ac.uk}
\author*[1]{A. Hillier}
\author*[1]{G. Murtas}
\author*[2]{G. J. J. Botha}
%\author*[1]{J. Mason}

\authormark{B. Snow \textit{et al.}}

\address[1]{\orgdiv{University of Exeter}, \orgname{Exeter}, \orgaddress{ \postcode{EX4 4PY}, \country{UK}}.,}
\address[2]{\orgdiv{Northumbria University}, \orgname{Newcastle upon Tyne}, \orgaddress{ \postcode{NE1 8ST}, \country{UK}}.,}

\received{}
\revised{}
\accepted{}

\abstract{
Compressible magnetohydrodynamic (MHD) turbulence is a common feature of astrophysical systems such as the solar atmosphere and interstellar medium. Such systems are rife with shock waves that can redistribute and dissipate energy.
For an MHD system, three broad categories of shocks exist (slow, fast or intermediate) however the occurrence rates of each shock type is not known for turbulent systems. 
Here we present a method for detecting and classifying the full range of MHD shocks applied to the Orszag-Tang vortex. Our results show that the system is dominated by fast and slow shocks, with far less frequent intermediate shocks appearing most readily near magnetic reconnection sites. We present a potential mechanism that could lead to the formation of intermediate shocks in MHD systems, and study the coherency and abundances of shocks in compressible MHD turbulence.
}


\end{Frontmatter}

\section{Introduction}
\label{sec:intro}

Many astrophysical systems are both highly compressible and turbulent, e.g., the solar atmosphere \citep{Ulmschneider1970,Carlsson1992,Houston2020,Reardon2008}  and interstellar medium \citep{Draine1983,Elmegreen2004}. A common feature of compressible turbulence is shocks, where the fluid properties change drastically over a small area. Such sharp structures are important for dissipation. For systems that are governed by magnetohydrodynamic (MHD) equations, there are three broad categories of stable shocks: slow, fast and intermediate \citep[e.g.,][]{Tidman1971}. Understanding the types of shocks that form in a system and where they are likely to form is critical to understand the energy transfer in a turbulent system.

Significant work has been performed to analyse the fast and slow shocks in turbulent systems \citep[e.g.,][]{Park2019,Komissarov2012,Orta2003}, however, intermediate shocks are far less studied
and their existence has been somewhat controversial in the past 
\citep[][]{Karimabadi1995,Wu1988}. 
These shock jumps are fully permitted by MHD equations \citep{Hau1989} and have recently been observed in the solar chromosphere \citep{Houston2020}. However, when and how intermediate shocks can form, and their relative stability, is far from understood. The `strong' intermediate shock (transition from super-Alfv\'enic to sub-slow) is smoothly connected to the slow mode shocks \citep{Hau1989}. In two-fluid simulations, it was shown that a slight upstream acceleration of the plasma can result in an intermediate shock forming \citep{Snow2019}. Previous studies have suggested that intermediate shocks may be related to magnetic reconnection \citep{Ugai1994,LaBelleHamer1994}.

Here we present a method to detect and classify the full range of MHD shocks from a 2D simulation and apply this to the Orszag-Tang vortex. The results indicate that the vast majority of the shocks are either fast or slow, and hence these shocks are responsible for the majority of the dynamic consequences. Less frequent intermediate shocks are detected and appear to form in high-current regions, that may arise due to turbulent inflow to a reconnecting region.

\section{Methods}

\subsection{Classification of shocks}

Shocks can be classified based on their upstream and downstream velocity relative to the characteristic speeds of the system, see Table \ref{tab:shocktrans}. For correct classification, the velocity must be in the shock frame, i.e., a frame of reference where the shock is stationary, for example, the de Hoffmann-Teller frame \citep{Hoffmann1950} where the electric field is zero either side of the shock. In the de Hoffmann-Teller frame, it is possible to derive an equation that gives all possible steady-state transitions as a function of the upstream (u) and downstream (d) Alfv\'en Mach numbers, and the upstream plasma-$\beta$ and angle of magnetic field \citep{Hau1989}.

\begin{table}[!h]
\vspace*{6pt}
\TBL{\caption{Possible stable MHD shock transitions based on the upstream and downstream states \label{tab:shocktrans}}}
{\begin{fntable}
%\begin{shaded}
\begin{tabular}{ccc}\toprule
    \textbf{Upstream state ($u$)} & \textbf{Downstream state ($d$)} & \textbf{Shock classification} \\ \midrule
    Super-fast (1) & Sub-fast (2)& $1\rightarrow 2$ Fast \\ \midrule
    Super-slow (3)& Sub-slow (4)& $3\rightarrow 4$ Slow \\ \midrule
    Super-fast (1)& Super-slow (3)& $1\rightarrow 3$ Intermediate \\ \midrule
    Super-fast (1)& Sub-slow (4)& $1\rightarrow 4$ Intermediate \\ \midrule
    Sub-fast (2)& Super-slow (3)& $2\rightarrow 3$ Intermediate \\ \midrule
    Sub-fast (2)& Sub-slow (4)& $2\rightarrow 4$ Intermediate \\ \botrule
\end{tabular}
%\end{shaded}
\footnotetext{\textit{Note:} Transitions of the form $u>d$ are entropy-forbidden so not listed here.}
%\footnotetext[1]{This is an example of table footnote}%
\end{fntable}}
\vspace*{-18pt}
\end{table}

\subsection{Identification technique}

To automatically identify shocks in our simulations, we use the following procedure:

\begin{enumerate}
    \item Identify shock candidates based on the divergence of the velocity field ($\nabla \cdot \textbf{v} < 0$ is a necessary condition for a shock). 
    \item Calculate the parallel and perpendicular unit vectors based on the maximum density gradient. The list of possible shocks is then filtered by checking the density gradient along a line perpendicular to the shock front. If the shock candidate is not a local maxima or minima of the perpendicular density gradient, then it is not the centre of the feature and is discarded. This prevents the numerical (and physical) finite width of the shock resulting in multiple detections.
    \item Estimate the shock frame from the steady-state conservation of mass equation. In the deHoffman-Teller shock frame, the conservation of mass equation becomes $\left[ \rho v_\perp \right]^u _d =0$. To transfer the simulation laboratory frame to the shock frame, we need the shock velocity $v_s$. By rewriting the mass conservation including this constant shock velocity as $\left[ \rho (v_\perp+v_s) \right]^u _d =0$, where $v_\perp$ is directly from the simulation, we can estimate the shock velocity by rearranging as $v_s=\frac{\rho ^d v_{\perp}^d-\rho ^u v_{\perp}^u}{\rho^u-\rho^d}$. 
    %\item Calculate the upstream and downstream slow, Alfv\'en and fast Mach numbers. 
    \item Compare the upstream and downstream Mach numbers to calculate the shock transition. 
\end{enumerate}

There are a number of assumptions that go into this identification technique, with the strongest being that the shocks are steady-state. 
In a highly dynamic simulation, this may prevent transient shocks being detected as readily and could lead to erroneous shock detection. 
As such, our identification procedure is likely an underestimate of the number of shocks in the system. Furthermore, we include additional checks to confirm the compressibility is greater than unity for all shocks, and that there is a reversal in the magnetic field across the interface for the intermediate shocks.

\begin{figure}
    \centering
    \begin{overpic}[width=0.32\linewidth,clip=true,trim=3.6cm 7.0cm 4.0cm 7.7cm]{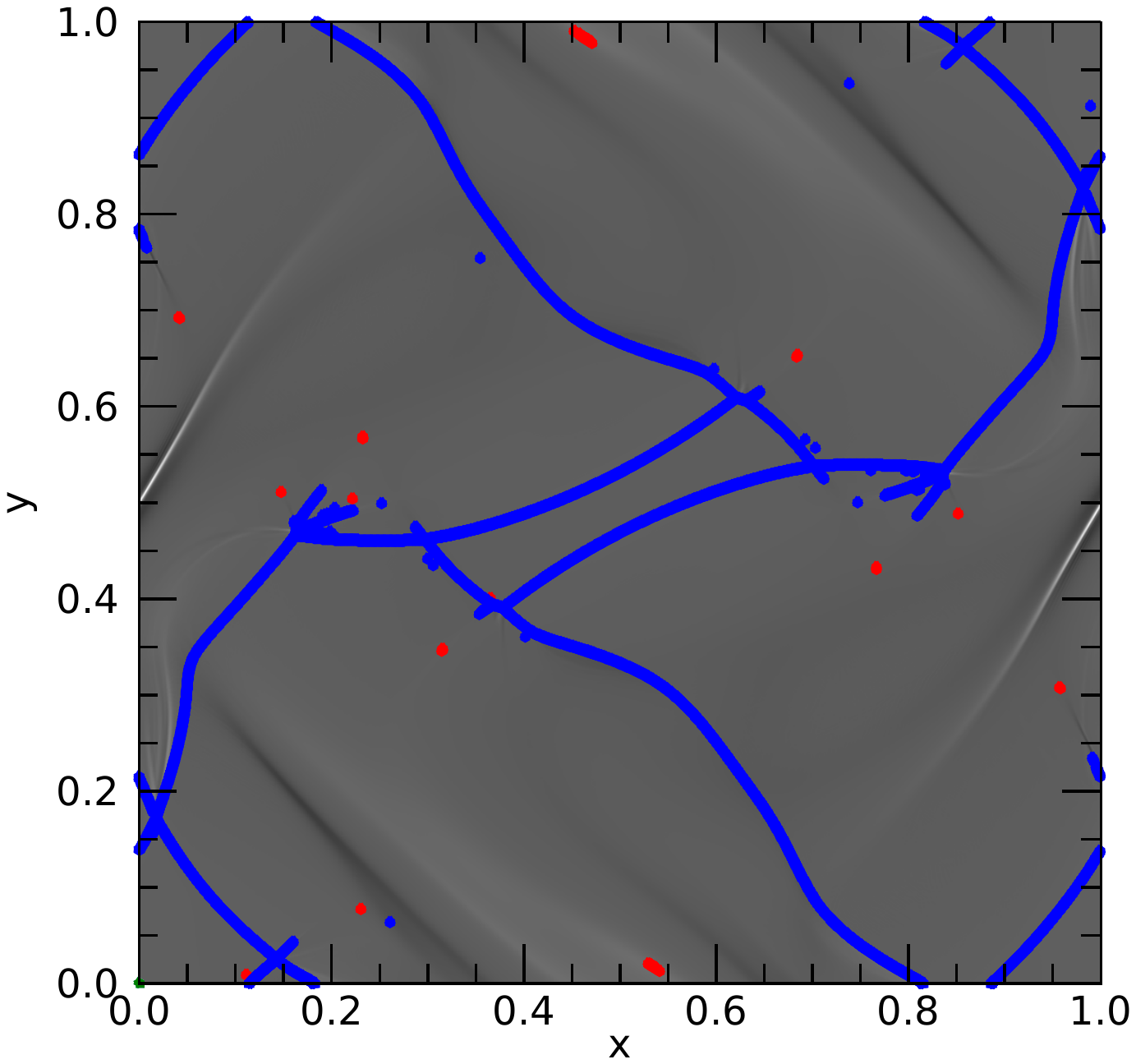}
     \put (18,80) {\textcolor{white}{\large(a)}}
    \end{overpic}
    \begin{overpic}[width=0.32\linewidth,clip=true,trim=3.6cm 7.0cm 4.0cm 7.7cm]{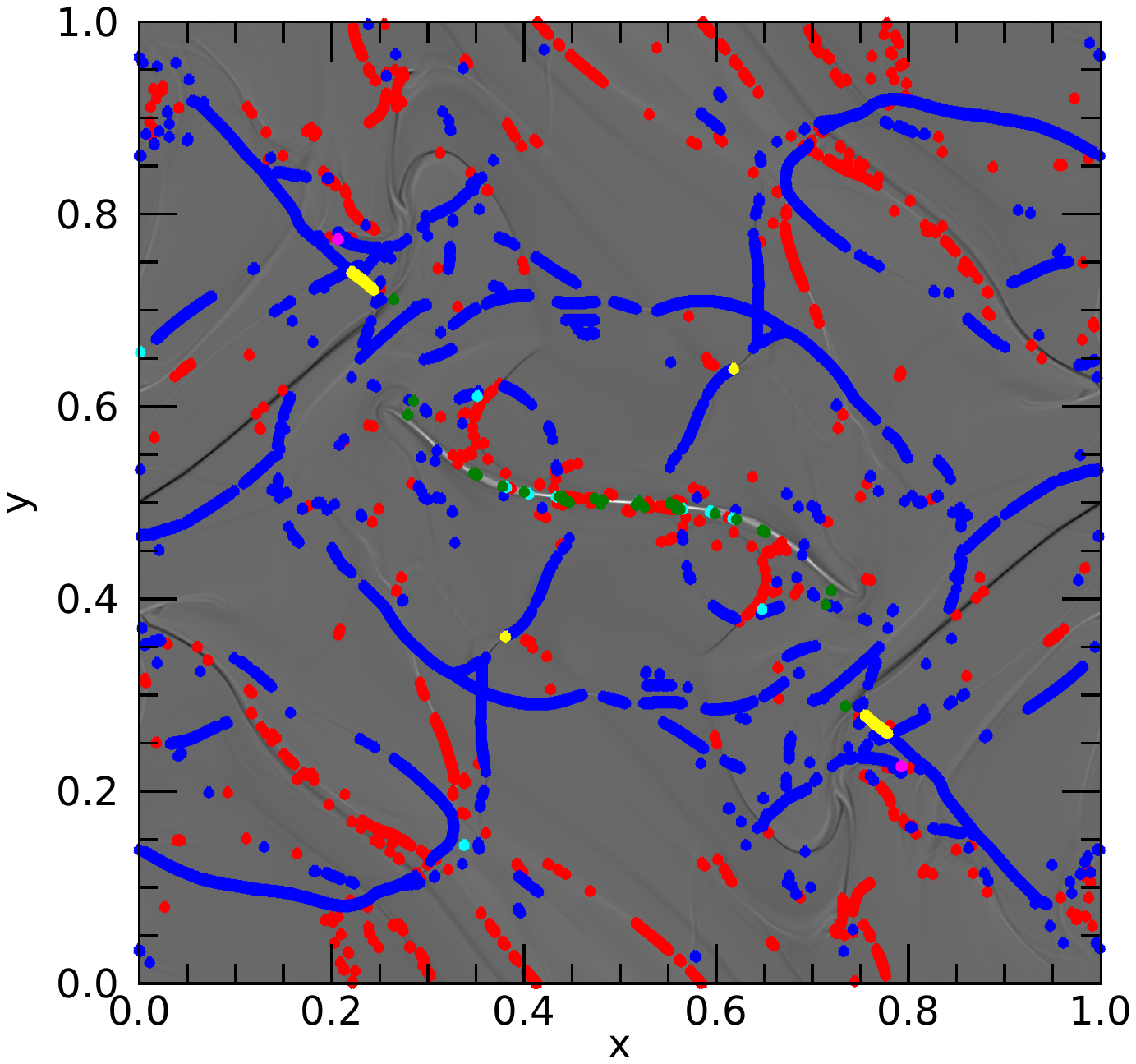} 
     \put (18,80) {\textcolor{white}{\large(b)}}
    \end{overpic}
    \begin{overpic}[width=0.32\linewidth,clip=true,trim=3.6cm 7.0cm 4.0cm 7.7cm]{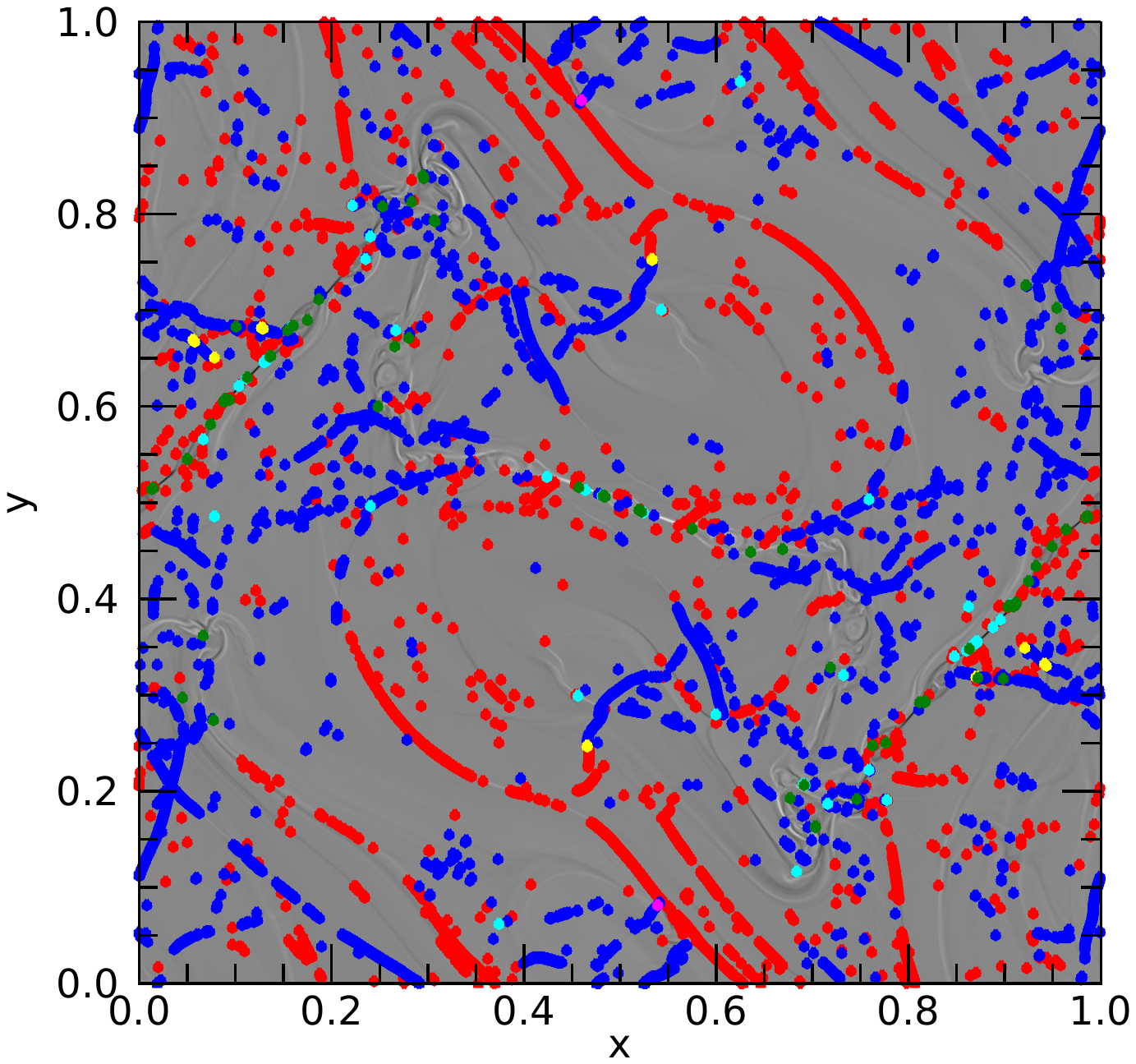}
     \put (18,80) {\textcolor{white}{\large(c)}}
    \end{overpic} \\
    \begin{overpic}[width=0.32\linewidth,clip=true,trim=3.6cm 7.0cm 4.0cm 7.7cm]{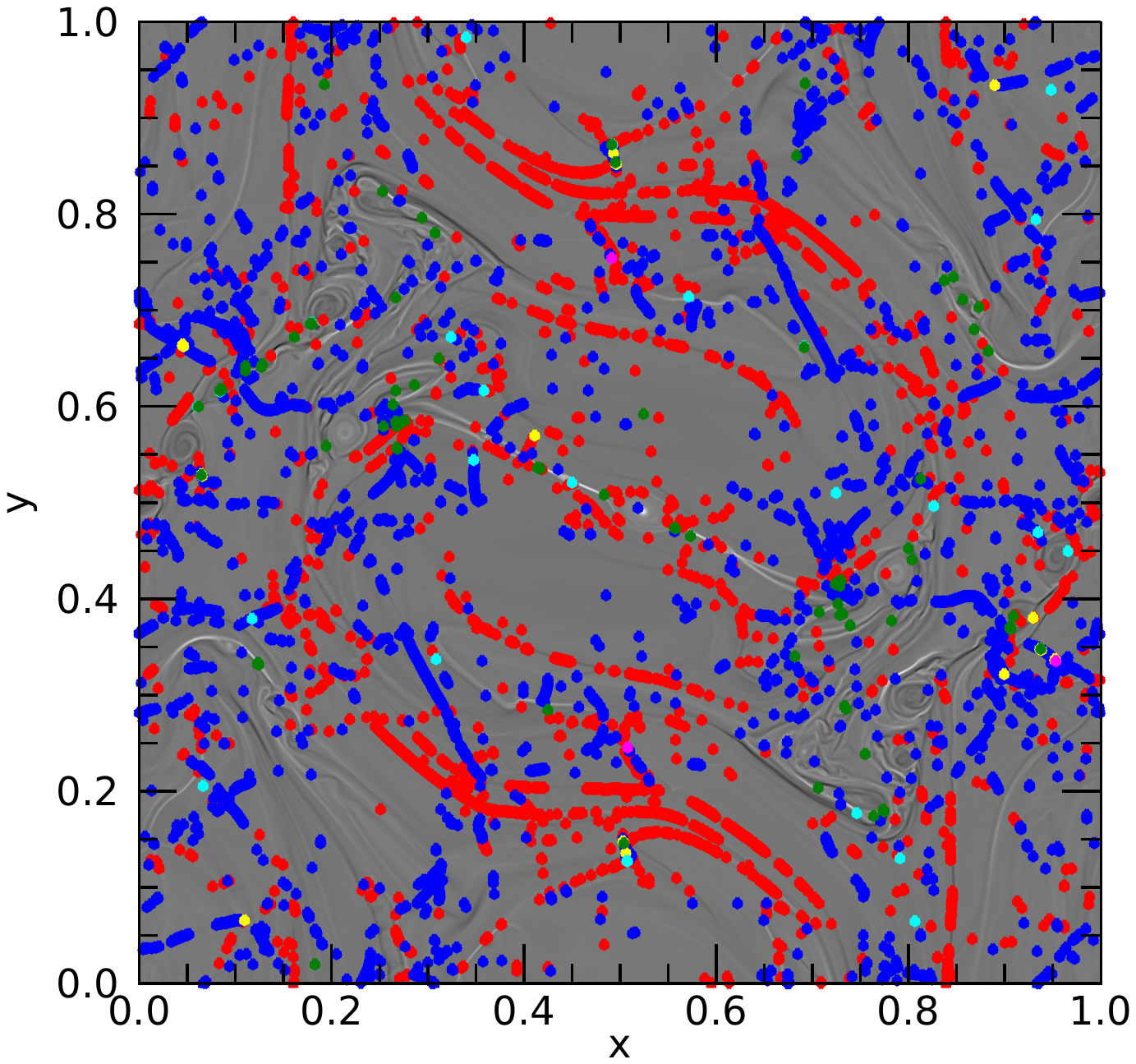}
     \put (18,80) {\textcolor{white}{\large(d)}}
    \end{overpic}
    \begin{overpic}[width=0.34\linewidth,clip=true,trim=1.0cm 5.5cm 1.3cm 6.0cm]{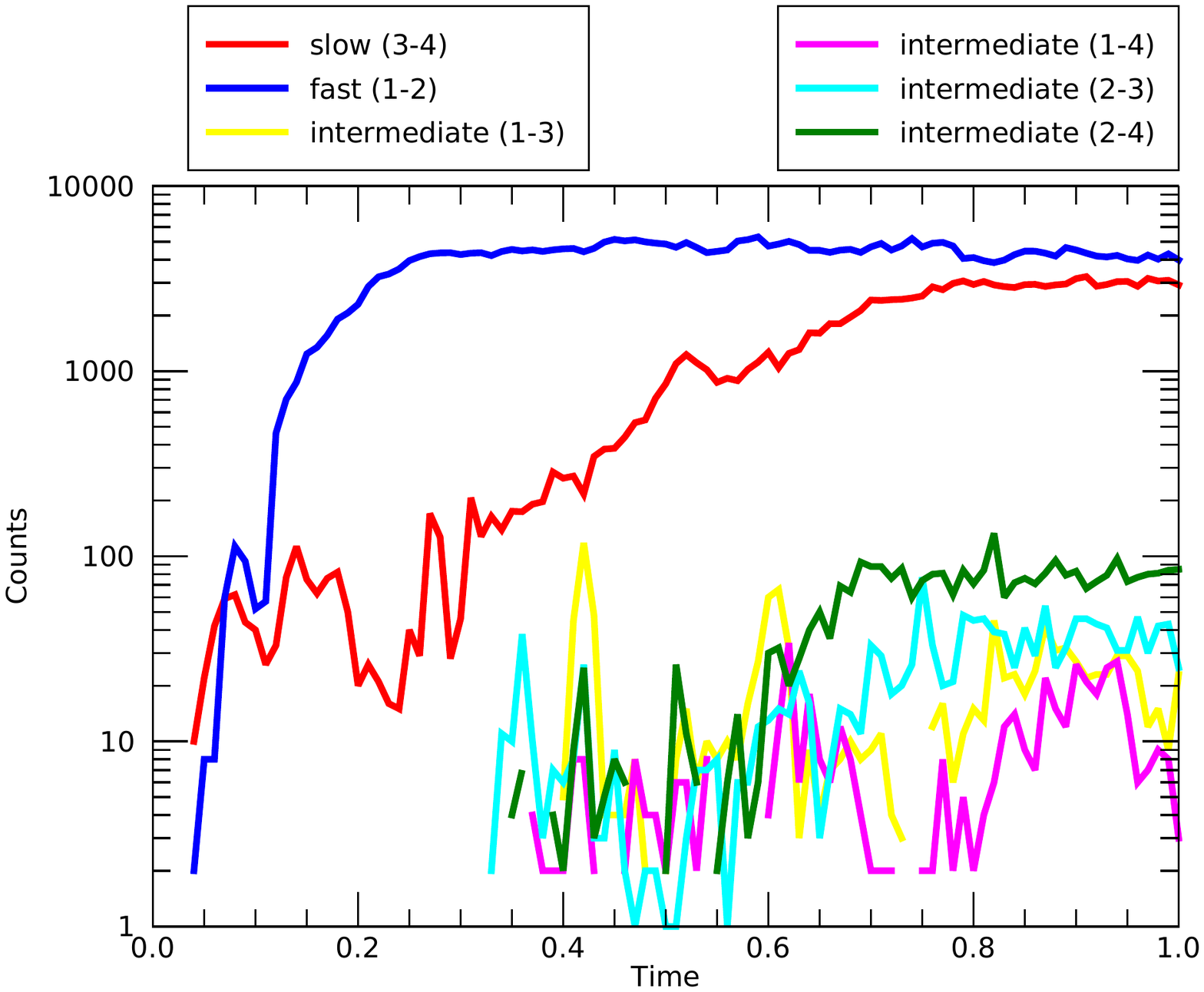}
     \put (18,60) {\large(e)}
    \end{overpic}
    \begin{overpic}[width=0.30\linewidth,clip=true,trim=2.0cm 6.8cm 3.6cm 7.3cm]{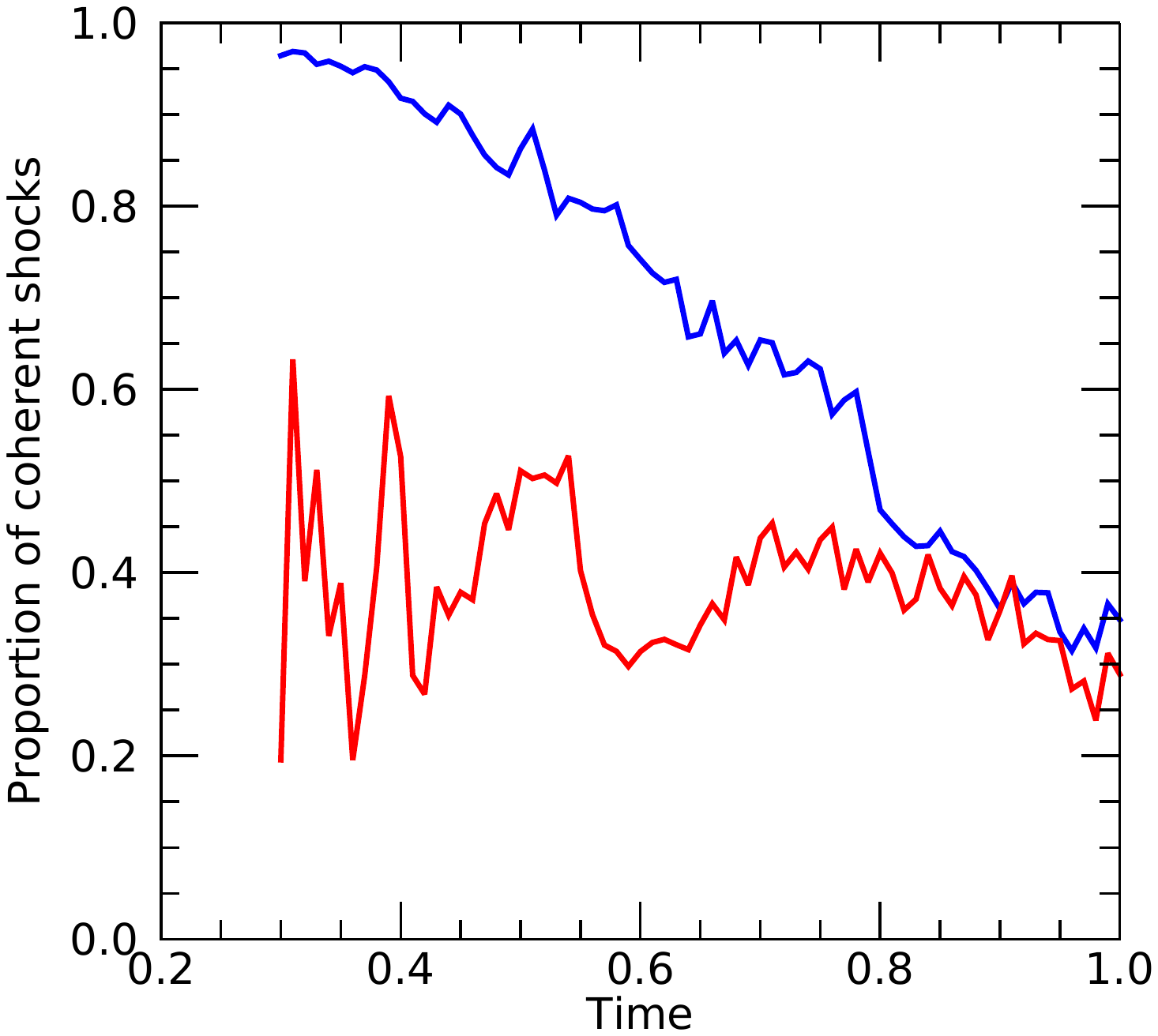}
     \put (22,70) {\large(f)}
    \end{overpic}
    \caption{Shocks detected in the Orszag-Tang vortex in MHD at $t=0.3 \mbox{(a)},0.6 \mbox{(b)},0.8 \mbox{(c)},1.0 \mbox{(d)}$. Background shows the out-of-plane current. Overplotted are the detected slow (red), fast (blue) and intermediate (yellow, magenta, cyan, green) shocks. The occurrence counts of each type of shock through time is shown in the panel (e). Panel (f) shows the coherency of fast and slow shocks through time}
    \label{fig:shockid}
\end{figure}

\subsection{MHD simulation}

To test this shock detection code, we use the Orszag-Tang (OT) shock vortex \citep{Orszag1979}. The OT initial conditions produce a decaying compressible turbulent system starting from the following initial conditions:
\vspace{-1.6cm}
\begin{multicols}{2}
\begin{gather}
    v_{x}=-v_0 \sin (2 \pi y), \\
    v_{y}= v_0 \sin (2 \pi x), \\
    B_{x}=-B_0 \sin (2 \pi y), \\
    B_{y}= B_0 \sin (4 \pi y), 
\end{gather}\break
\begin{gather}
    \rho = 25/(36 \pi), \\
    P=5/(12 \pi), \\
    v_0=1, \\
    B_0=1/\sqrt{4 \pi},
\end{gather}
\end{multicols}
\vspace{-0.2cm}
\noindent for density $\rho$, pressure $P$, velocity $\textbf{v}=[v_x,v_y,0]$ and magnetic field $\textbf{B}=[B_x,B_y,0]$. 
The Orszag-Tang vortex has been well studied in the literature \citep{Dahlburg1989,Picone1991,Jiang1999,Parashar2009,Uritsky2010}.
%In this paper, the system is 2D, however 3D OT vortexes have been studied in other works REFERENCES. 
The initial conditions are evolved in 2D for ideal MHD equations using the 4th order central-difference solver in the (P\underline{I}P) code \citep{Hillier2016}, which has been used previously to study shocks \citep[e.g.,][]{Snow2021}. The simulations are performed using $1024\times1024$ cells with periodic boundary conditions. 
%\textcolor{red}{The 4th order solver has been used previously to study shocks with the (P\underline{I}P) code \citep[e.g.,][]{Snow2021}.}

\subsection{Results}

Figure \ref{fig:shockid} shows the evolution of the Orszag-Tang vortex through time, with the identified shock pixels coloured according to their classification.
The simulation shows the full range of shock transitions exists in our simulation, with the most prevalent being the fast and slow mode shocks. Intermediate shocks are detected in the simulation however these are far less frequent and appear later than the fast or slow shocks. 

Figure \ref{fig:shockid}e shows the number and type of detected shocks in the simulation through time. At very early times in the simulation ($t<0.15$), the detections of shocks is sporadic as the system is evolving rapidly and the assumption of a steady-state shock is not valid. After $t=0.15$, large-scale fast-mode shocks are generated due to the initial conditions of the Orszag-Tang vortex.
 As the system develops and the shocks interact, slow-mode shocks become increasingly detected in the system, and fast and slow shocks have comparable detection counts near the end of the simulation ($t=1$). The relative counts of fast and slow mode shocks is likely dependent on the initial conditions of the system \citep{Dahlburg1989,Picone1991}.

A metric to determine the coherency of shocks can be calculated by measuring the number of similar shocks within a given radius. Here we use a radius of three grid cells and classify any shocks that have more than six nearby shocks of the same type as coherent. Figure \ref{fig:shockid}f shows the proportion of coherent fast and slow shocks through time. At early times in the simulation, the system is dominated by long, continuous fast-mode shocks, and hence the proportion of coherent shocks is very high. As the simulation evolves and the fast-mode shocks interact and break up, the coherency decreases. For the slow-mode shocks, the coherency is fairly constant through time at around 40\%. From Figure \ref{fig:shockid}b,c,d one can see that the the large scale slow-mode shocks may exist but are not detected as continuous features, possibly due to either limitations of the detection method.   

It has been suggested that the corrugation instability \citep[e.g.,][]{Stone1995} may result in far fewer slow-mode shocks being detectable in turbulent simulations \citep{Park2019}. However, here there are comparable numbers of fast and slow shocks in the system after $t=0.8$, and large un-corrugated coherent slow-mode shock structures exist in the simulation, see Figure \ref{fig:shockid}b-d. A recent study has shown that the the corrugation instability can actually increase the number of detected shock pixels due to an increased shock length \citep{Snow2021}. 

\subsubsection{Potential mechanism for the formation of intermediate shocks}

\begin{figure}
    \centering
    \includegraphics[width=0.95\textwidth,clip=true,trim=0.0cm 0.0cm 0.0cm 0.0cm]{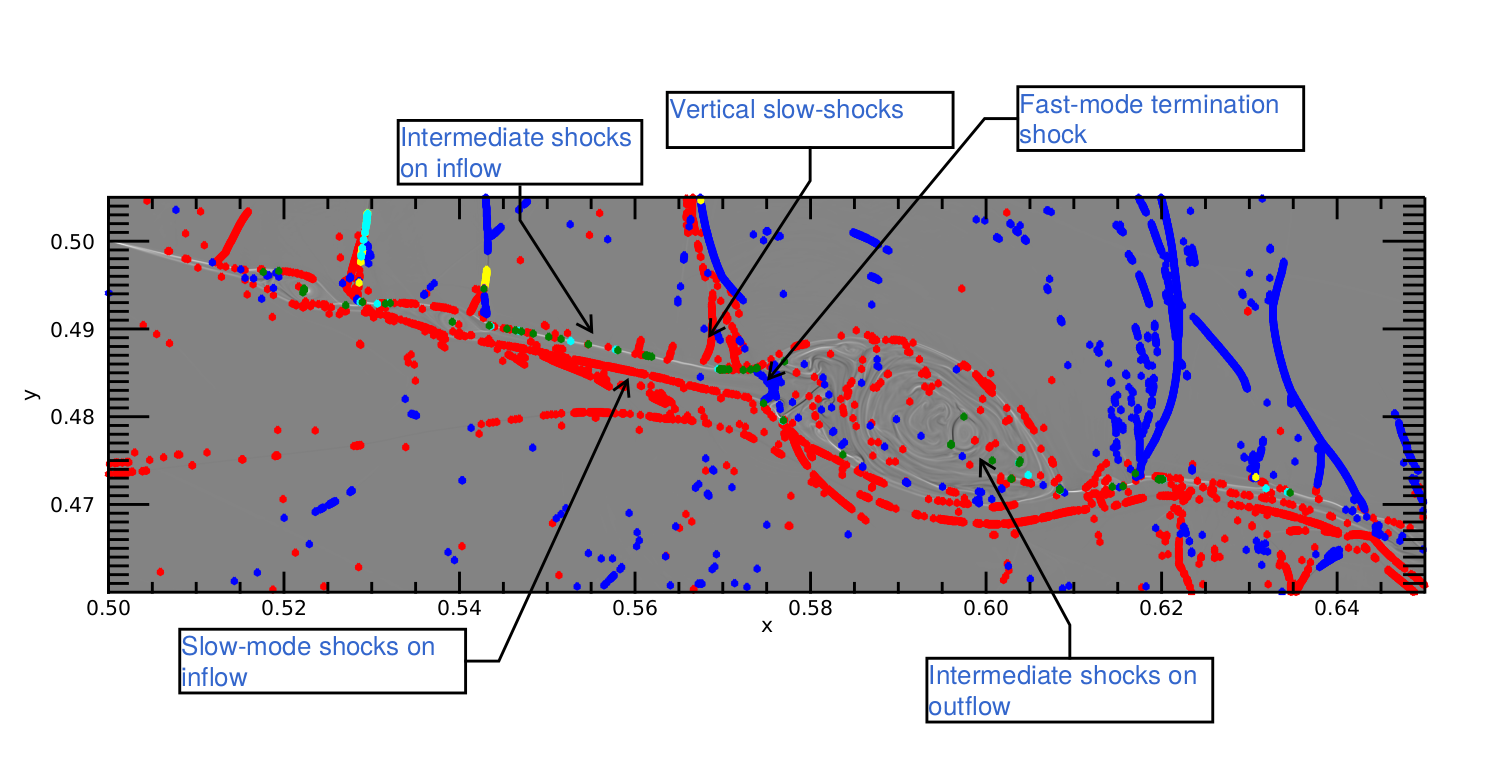}
    \caption{Plasmoid with intermediate shocks forming on inflow region. Fundamental structures of the plasmoid are annotated}
    \label{fig:plasmoid}
\end{figure}

The intermediate shocks appear to form near and around high-current regions. As such, one may consider the link between magnetic reconnection and intermediate shocks in turbulent systems. Note that not all high current regions feature intermediate shocks, but intermediate shocks are usually located at a high-current region.
For fast magnetic reconnection, one may expect switch-off slow-mode shocks \citep[e.g.,][]{Petschek1964}, where the inflow Alfv\'en mach number is unity. High resolution MHD simulations have shown that switch-off slow-mode shocks are integral to plasmoid formation \citep{Zenitani2011}. Now, since a switch-off slow-mode shock has an inflow Alfv\'en Mach number of one, it is on the cusp of being a $2\rightarrow4$ intermediate shock \citep[where the inflow is super-Alfv\'enic and the downstream is sub-slow, e.g., Figure 1 in][]{Hau1989}. As such, for a turbulent reconnection region, one would expect variations in the inflow quantities that could result in a super-Alfv\'enic inflow and an intermediate shock. This appears to be the formation mechanism occurring here. Specifically isolating a plasmoid in a high resolution MHD simulation, one sees remarkable similarities to the plasmoid schematic in \cite{Zenitani2011}, however, here intermediate shocks form on the inflow, see Figure \ref{fig:plasmoid}. In this example figure, the inflow to the plasmoid is asymmetric and slow-mode shocks form on the lower inflow, whereas intermediate shocks form on the upper inflow.

\begin{figure}
    \centering
    \begin{overpic}[width=0.95\textwidth,clip=true,trim=1.0cm 9.5cm 2.0cm 10.0cm]{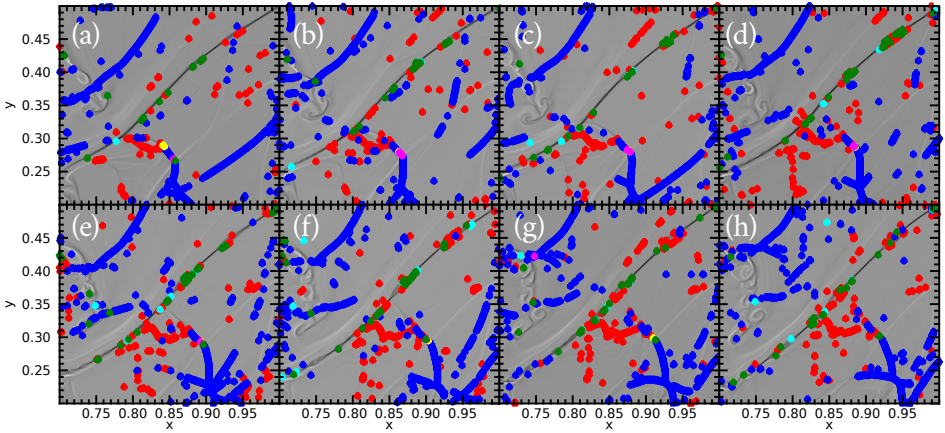}
    \put (8,40) {\textcolor{white}{\large(a)}}
    \put (30,40) {\textcolor{white}{\large(b)}}
    \put (52,40) {\textcolor{white}{\large(c)}}
    \put (73,40) {\textcolor{white}{\large(d)}}
    \put (8,21) {\textcolor{white}{\large(e)}}
    \put (30,21) {\textcolor{white}{\large(f)}}
    \put (52,21) {\textcolor{white}{\large(g)}}
    \put (73,21) {\textcolor{white}{\large(h)}}
    \end{overpic}
    \caption{Snapshots of a reconnecting region in the time frame $t=0.66-0.73$ (from (a)-(h)) at intervals of $0.01$. Background colour is the out-of-plane current density}
    \label{fig:recregion}
\end{figure}

In terms of relative stability, in these simulations the intermediate shocks can be regularly detected across a few times steps, and by eye one can track certain shocks across outputs, however, others are less stable, see Figure \ref{fig:recregion}. Through time, the intermediate shocks are routinely detected at the high-current regions. Further work is needed to analyse the stability and evolution of intermediate shocks. 

\subsection{Consistency of results at higher grid resolutions}

A high resolution ($16384^2$) simulation of the Orszag-Tang vortex was performed to analyse the consistency of the results. At time $t=1$ the number of pixels that satisfy a shock jump condition in the high-resolution simulation are: 160819 (slow), 316029 (fast), 350 (intermediate 1-2), 23 (intermediate 1-4), 836 (intermediate 2-3), and 1394 (intermediate 2-4). The increased grid resolution has lead to many more shocks being detected, however, proportionally the results are similar. There is a factor of $\approx 2$ more fast than slow shocks (as in the $1024^2$ simulation, see Figure \ref{fig:shockid}). Intermediate shocks are proportionally less frequent; intermediate 2-4 shocks are $\approx$ 3\% as frequent as slow shocks at 1024$^2$, and $\approx$ 0.1\% in the high-resolution simulation. This may be due to the larger scale shocks contributing more to the overall shock count and requires further study. However, fast and slow shocks are most abundant, and the intermediate shocks form at high-current regions in both simulations indicating that these are consistent results.

\section{Conclusion}

The developed method for identifying shocks in MHD systems detects the full range of MHD shocks in the Orszag-Tang vortex. The MHD simulation is dominated by fast and slow shocks implying that the vast magority of the shock-related dissipation occurs due to fast and slow shocks. Furthermore, the fast and slow shocks occur with roughly equal detected pixel counts at $t=1$ implying that the corrugation instability is not preventing slow-mode shocks from being detected. Less-frequent intermediate transitions are also detected in the simulation that appear to be connected to high-current regions. 
%The intermediate shocks are typically located at the reconnection sites. 
In particular, the 2-4 intermediate shock may be formed due to fluctuations of the inflow velocity from a reconnection generated switch-off shock. Further work is needed to analyse the evolution and stability of intermediate shocks occurring in turbulent MHD systems.

\begin{Backmatter}

\paragraph{Acknowledgements} 
We would like to thank the stimulating discussions with Joanne Mason, Pierre Lesaffre, Andrew Lehmann and Thibaud Richard.

\paragraph{Author Contributions}
BS designed the study and wrote the manuscript. AH, GJJB provided insight and assisted in understanding the results. GM assisted in benchmarking the shock identification code.

\paragraph{Funding Information}
BS and AH are supported by STFC research grant ST/R000891/1 and ST/V000659/1
%\textcolor{red}{AH also acknowledges support by STFC Ernest Rutherford Fellowship grant number ST/L00397X/2}

\paragraph{Data Availability Statement}
The simulation data from this study are available from BS upon reasonable request. The (P\underline{I}P) code is available at \href{https://github.com/AstroSnow/PIP}{https://github.com/AstroSnow/PIP}. The IDL scripts used for this manuscript are available at \href{https://github.com/AstroSnow/Orszag-Tang-Shock-Detection}{https://github.com/AstroSnow/Orszag-Tang-Shock-Detection}

\paragraph{Conflict of Interest}
BS, AH, GM, GJJB have no conflicts of interest to declare.

\bibliographystyle{apalike}
\bibliography{shockidrefs}

\end{Backmatter}

% \section{Appendix}

\end{document}